\documentclass[twocolumn,10pt]{asme2e}
\usepackage{amssymb}
\usepackage[pdftex]{graphicx}
\usepackage{amsmath}
\usepackage{graphicx}
\usepackage{caption}
\usepackage{subcaption}
\usepackage{color}
\usepackage{colortbl}
\usepackage{setspace}
\usepackage{titlesec}
\titlespacing*{\section}{0pt}{5pt}{5pt}
\expandafter\def\expandafter\normalsize\expandafter{%
	\normalsize
	\setlength\abovedisplayskip{8pt}
	\setlength\belowdisplayskip{8pt}
	\setlength\abovedisplayshortskip{10pt}
	\setlength\belowdisplayshortskip{10pt}
}
\graphicspath{{./Figures/}}

\definecolor{lightgray}{gray}{0.9}

\confshortname{ISPS 2016}
\conffullname{the ASME 2016 Conference on Information Storage and Processing Systems}
\confdate{June 20-21}
\confyear{2016}
\confcity{Santa Clara}
\confcountry{USA}
\title{DSP Implementation of a Direct Adaptive Feedfoward Control Algorithm for Rejecting Repeatable Runout in Hard Disk Drives}

\author{Jinwen Pan
    \affiliation{
	Department of Mechanical Engineering\\
	University of California, Berkeley\\
	Berkeley, California 94720\\
    Email: jinwen@berkeley.edu
    }	
}
\author{Prateek Shah 
	\affiliation{
	Department of Mechanical Engineering\\
	University of California, Berkeley\\
	Berkeley, California 94720\\
	Email: prateekshah@berkeley.edu
    }
}

\author{Roberto Horowitz 
    \affiliation{
    Department of Mechanical Engineering\\
	University of California, Berkeley\\
	Berkeley, California 94720\\
	Email: horowitz@berkeley.edu
    }
}

\begin{document}

\maketitle 
\def\Hi{$\mathcal{H}_\infty$ }
\def\H2{$\mathcal{H}_2$ }
\begin{abstract}\footnotetext{Submitted to ASME 2016 Conference on Information Storage and Processing Systems}
{\it A direct adaptive feedforward control method for tracking repeatable runout (RRO) in bit patterned media recording (BPMR) hard disk drives (HDD) is proposed. The technique estimates the system parameters and the residual RRO simultaneously and constructs a feedforward signal based on a known regressor. An improved version of the proposed algorithm to avoid matrix inversion and reduce computation complexity is given. Results for both MATLAB simulation and digital signal processor (DSP) implementation are provided to verify the effectiveness of the proposed algorithm.}
\end{abstract}
%

\section{INTRODUCTION}\label{sec:intro}
\begin{figure}[h!]
\centering
\includegraphics[width=0.8\columnwidth]{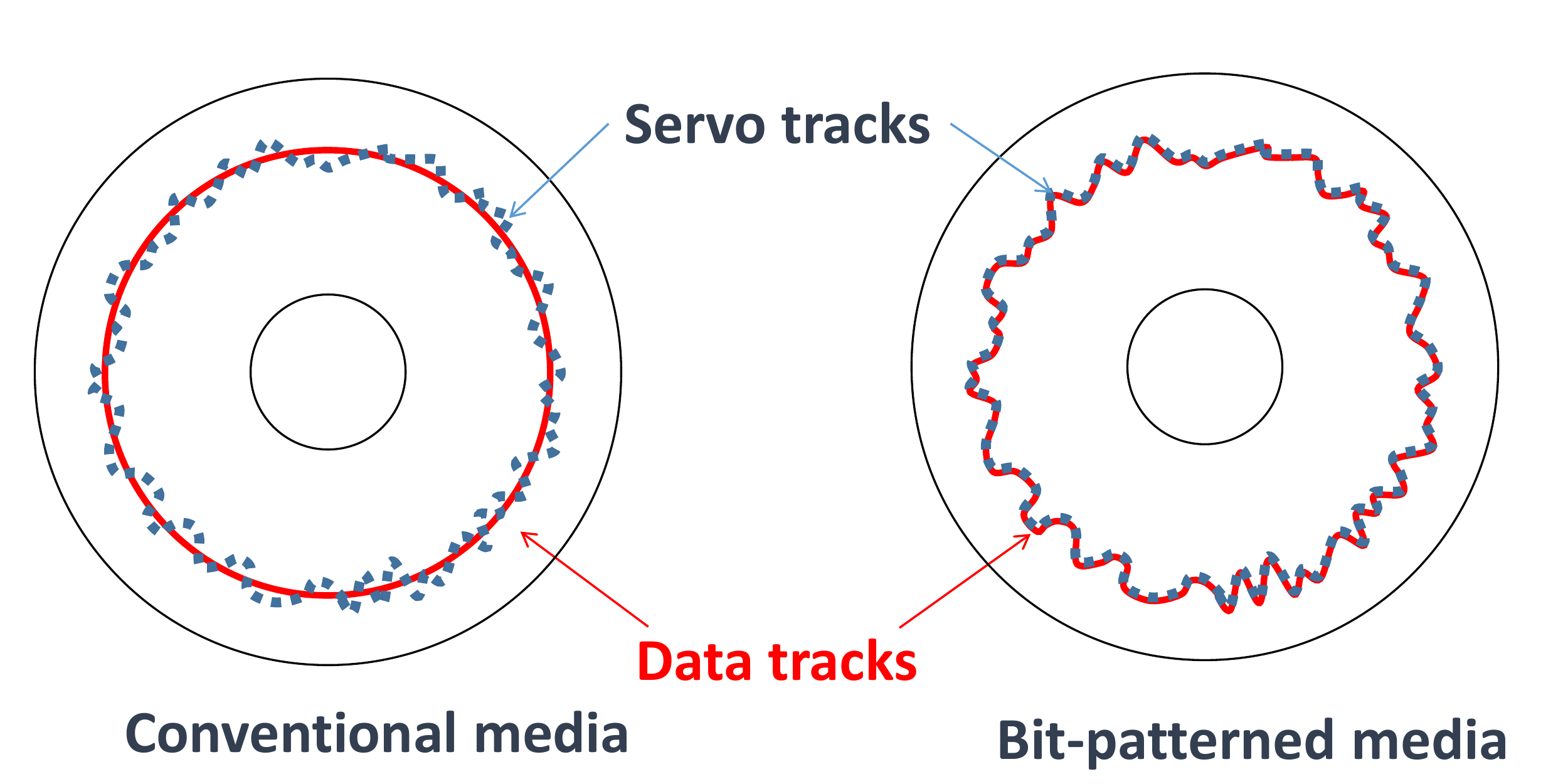}
\caption{Servo Track (dotted blue) and data track (solid red) in conventional and bit-patterned media}
\label{Fig:Conventional_BPM}
\end{figure}

Data bits are ideally written on concentric circular tracks in conventional HDDs that use magnetic disks with continuous media. This process is different in bit patterned media recording since data should be written on tracks with predetermined shapes, which are created by lithography on the disk \cite{shahsavari2015adaptive,keikha2013probabilistic}. 
As shown in Fig.~\ref{Fig:Conventional_BPM} the trajectories that are required to be followed by the servo system in BPMR are servo tracks, which are characterized by the servo sectors written on the disk. Deviation of a servo track from an ideal circular shape is called RRO. Therefore, the servo controller in BPMR has to follow the RRO which is unknown in the time of design, and as a result the servo control methodologies used for conventional drives \cite{kempf1993comparison} cannot be applied to BPMR directly. In our prior works, we proposed indirect adaptive control methods for mechatronic devices to compensate for unknown disturbances (such as RRO) \cite{shahsavari2014repeatable,shahsavari2014adaptive,shahsavari2016adaptive} and dynamics mismatches \cite{zhang2014adaptive,zhang2014adaptive2}.
In this paper, we propose a direct adaptive control method to address 
challenges specific to BPMR which are briefly listed here: (1) RRO profile is unknown; (2) RRO frequency spectrum can spread beyond the bandwidth of servo system, therefore, it will be amplified by the feedback controller; (3) RRO spectrum contains many harmonics of the spindle frequency (e.g. $\sim200$ harmonics) that should be attenuated, which increases the computational burden in the controller; (4) RRO profile is changing from track to track (i.e. it is varying on the radial direction);
(5) HDD servo dynamics changes from drive to drive and by temperature \cite{bagherieh2015online}.

The remainder of this paper is organized as follows. Section~\ref{sec:control} presents our direct adaptive feedforward control algorithm and section~\ref{sec:experiment} shows the real time DSP implementation results.

\section{Control Design}\label{sec:control}
\begin{figure}[t!]
\centering
\includegraphics[width=0.6\columnwidth]{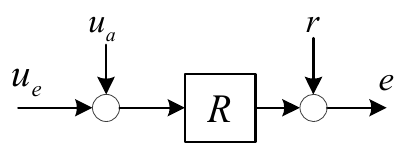}
\caption{Control architecture.}
\label{Fig:Control}
\end{figure}

The architecture that is considered for the servo control system is shown in Fig.~\ref{Fig:Control}. An add-on feedforward controller is designed for HDD. Without loss of generality, we chose VCM as an example. Here, $ R $ is the transfer function from VCM input to PES, $ u_e $ is an exogenous excitation signal, $ u_a $ is the feedforward signal, $ r $ is the unknown RRO with known frequencies and $ e $ is the PES. We aim to design an adaptive controller that generates $ u_a $ in order to fade the frequency contents of the error signal $ e $ at selective frequencies which correspond to the harmonics of spindle frequency (120Hz and its harmonics in our case).

\subsection{Basic Direct Adaptive Feedforward Control}\label{sec:algorithmB}
From Fig.~\ref{Fig:Control}, the PES can be written as
\begin{equation*}
e\left( k \right) = R\left( {{q^{ - 1}}} \right)\left( {{u_e}\left( k \right) + {u_a}\left( k \right)} \right) + r\left( k \right),
\end{equation*}
where $ R\left( {{q^{ - 1}}} \right) = B\left( {{q^{ - 1}}} \right)/A\left( {{q^{ - 1}}} \right) $ and can be expanded as
\begin{equation}\label{eq:ekab}
e\left( k \right) = {A^*}\left( {{q^{ - 1}}} \right)e\left( k \right) + B\left( {{q^{ - 1}}} \right){u_e}\left( k \right) + \theta _M^T{\phi _r}\left( k \right),
\end{equation}
where $ A\left( {{q^{ - 1}}} \right) = 1 - {A^*}\left( {{q^{ - 1}}} \right) $, $ {A^*}\left( {{q^{ - 1}}} \right) = {a_1}{q^{ - 1}} +  \cdots  + {a_{{n_a}}}{q^{ - {n_a}}} $, $ B\left( {{q^{ - 1}}} \right) = {b_1}{q^{ - 1}} +  \cdots  + {b_{{n_b}}}{q^{ - {n_b}}} $ and the residual error
\begin{equation}\label{eq:residual}
\theta _M^T{\phi _r}\left( k \right) = B\left( {{q^{ - 1}}} \right){u_a}\left( k \right) + A\left( {{q^{ - 1}}} \right)r\left( k \right),
\end{equation}
where $ {\phi _r}\left( k \right) $ is the regressor for RRO with known frequencies. In regressor form, PES is
\begin{equation*}
e\left( k \right) = \theta _A^T{\phi _e}\left( k \right) + \theta _B^T{\phi _{ue}}\left( k \right) + \theta _M^T{\phi _r}\left( k \right),	
\end{equation*}
and its estimation
\begin{equation*}
\hat e\left( k \right) = \hat \theta _A^T\left( k \right){\phi _e}\left( k \right) + \hat \theta _B^T\left( k \right){\phi _{ue}}\left( k \right) + \hat \theta _M^T\left( k \right){\phi _r}\left( k \right),	
\end{equation*}
where $ \theta _A^T = \left[ {{a_1},{a_2}, \cdots ,{a_{{n_a}}}} \right] $, $ {\phi _e}\left( k \right) $, $ \theta _B^T = \left[ {{b_1},{b_2}, \cdots ,{b_{{n_b}}}} \right] $ and $ {\phi _{ue}}\left( k \right) $ are regressors for $ e(k) $ and $ u_e(k) $. $ \hat \theta _A $, $ \hat \theta _B $ and $ \hat \theta _M $ are the estimates of $ \theta _A $, $ \theta _B $ and $ \theta _M $ and the updating law is
\begin{equation}\label{eq:updateABM}
\hat \theta \left( {k + 1} \right) = \hat \theta \left( k \right) + K\left( k \right)\phi \left( k \right)\tilde e\left( k \right)/\left( {1 + {\phi ^T}\left( k \right)\phi \left( k \right)} \right),	
\end{equation}
where $ \tilde e\left( k \right) = e\left( k \right) - \hat e\left( k \right) $, $ \hat \theta \left( k \right) = {\left[ {\hat \theta _A^T\left( k \right),\hat \theta _B^T\left( k \right),\hat \theta _M^T\left( k \right)} \right]^T} $, $ \phi \left( k \right) = {\left[ {\phi _e^T\left( k \right),\phi _{ue}^T\left( k \right),\phi _r^T\left( k \right)} \right]^T} $ and $ K(k) $ is a decreasing gain. Eq.~(\ref{eq:updateABM}) indicates that both of the system and the residual RRO are estimated simultaneously. The feedforward control signal is constructed using the same regressor as the RRO yielding
\begin{equation}\label{eq:ffwd}
{u_a}\left( k \right) = \hat \theta _D^T\left( k \right){\phi _r}\left( k \right).	
\end{equation}
In Eq. (\ref{eq:residual}), using $ \hat \theta _M (k) $ and $ \hat B\left( {{q^{ - 1}}} \right) $ instead of $ \theta _M $ and $ B(q^{-1}) $, approximately we have
\begin{equation}\label{eq:appxieps}
\hat B\left( {{q^{ - 1}}} \right)\left[ {\hat \theta _D^T\left( k \right){\phi _r}\left( k \right)} \right] + A\left( {{q^{ - 1}}} \right)r\left( k \right) = \hat \theta _M^T\left( k \right){\phi _r}\left( k \right),
\end{equation}
where $ \hat B\left( {{q^{ - 1}}} \right) = {{\hat b}_1}\left( k \right){q^{ - 1}} +  \cdots  + {{\hat b}_{{{\hat n}_b}}}\left( k \right){q^{ - {{\hat n}_b}}} $ with $ \hat n_b $ the estimate of $ n_b $. Since $ A\left( {{q^{ - 1}}} \right)r\left( k \right) = \theta _{\bar R}^T{\phi _r}\left( k \right) $ with $ {\theta _{\bar R}} $ an unknown vector and $ {\phi _r}\left( k \right) $ a nonzero vector, from Eq.~(\ref{eq:appxieps}) we have
\begin{equation*}
{{\hat \theta }_D}\left( k \right) + {\theta _{\bar{R}}}(k) = D_{\hat B}^{ - 1}\left( k \right){{\hat \theta }_M}\left( k \right),
\end{equation*}
where $ {D_{\hat B}}\left( k \right) $ can be formed based on the magnitude and phase of $ \hat B\left( {{e^{ - j{\omega _i}}}} \right) $, $ i = 1,2, \cdots ,{n_r} $ with $ n_r $ the number of frequencies to cancel. The updating law for $ {{\hat \theta }_D}\left( k \right) $ is 
\begin{equation}\label{eq:errorThm}
{{\hat \theta }_D}\left( {k + 1} \right) = {{\hat \theta }_D}\left( k \right) - \alpha D_{\hat B}^{ - 1}\left( k \right){{\hat \theta }_M}\left( k \right).
\end{equation}
The inverse of $ {D_{\hat B}}\left( k \right) $ involves inverting the estimated magnitudes that might be very small in transition, especially when $ {\hat \theta_B} $ is initialized by zeros. In that case, any small fluctuation of $ {\hat \theta_B} $ can cause large transient error. A smoothing on magnitude and phase of $ \hat B\left( {{e^{ - j{\omega _i}}}} \right) $ has to be designed to relax transient errors. The basic direct adaptive feedforward control algorithm is summarized in Table.~\ref{tab:controlalgorithm}.

\begin{spacing}{1.1}
	\begin{center}
		\begingroup\setlength{\fboxsep}{0pt}
		\colorbox{lightgray}{%
			\begin{tabular}{|l|}
				\hline 
				0. Initialize the regressors $ {\phi _e}\left( k \right) $,  $ {\phi _{ue}}\left( k \right) $ and $ {\phi _r}\left( k \right) $;
				\\
				1. Apply $ u_e(k) $ and $ u_a(k) $ to VCM;
				\\
				2. Subtract $ \hat{e}(k) $ from PES to determine the estimate error $ \tilde{e}(k) $;
				\\
				3. Update the parameters $ \hat \theta (k) $ using Eq.~(\ref{eq:updateABM});
				\\
				4. Update the matrix  $ {D_{\hat B}}\left( k \right) $ from $ \hat \theta_B(k) $ and compute its inverse;
				\\
				5. Update $ \hat{\theta}_D(k) $ using Eq.~(\ref{eq:errorThm}) and compute $ u_a(k) $ from Eq.~(\ref{eq:ffwd}).
				\\
				\hline				
			\end{tabular}%
		}\endgroup
		\captionof{table}{Basic Direct Adaptive Feedforward Control}\label{tab:controlalgorithm}
	\end{center}
\end{spacing}

\subsection{Improved Direct Adaptive Feedforward Control}\label{sec:algorithmI}
As mentioned earlier, computational complexity of inverting $ {D_{\hat B}}\left( k \right) $ grows as the number of frequencies increases which is a crucial burden in DSP implementation. In this section, we will provide an improved version to avoid matrix inversion. By applying ``swapping lemma'' to \eqref{eq:appxieps}, we have

\begin{equation}\label{eq:swap}
\hat \theta _D^T\left( k \right)\left[ {\hat B\left( {{q^{ - 1}}} \right){\phi _r}\left( k \right)} \right] + A\left( {{q^{ - 1}}} \right)r\left( k \right) = \hat \theta _M^T\left( k \right){\phi _r}\left( k \right),
\end{equation}
therefore, the updating law for $ {{\hat \theta }_D}\left( k \right) $ is
\begin{equation}\label{eq:errorNew}
{{\hat \theta }_D}\left( {k + 1} \right) = {{\hat \theta }_D}\left( k \right) - \alpha \left[ {\hat B\left( {{q^{ - 1}}} \right){\phi _r}\left( k \right)} \right]\left[ {\hat \theta _M^T\left( k \right){\phi _r}\left( k \right)} \right].
\end{equation}
Note that in \eqref{eq:errorNew} no matrix inverse is required. The improved direct adaptive feedforward control algorithm is summarized in Table.~\ref{tab:updatealgorithm} where the first three steps are the same as those in Table.~\ref{tab:controlalgorithm}. To be noted here, the proposed direct adaptive feedforward control algorithm and its improved version can be directly extended to the second-stage actuator which is mili-actuator (MA) responsible for high frequency RRO.

\begin{spacing}{1.1}
	\begin{center}
		\begingroup\setlength{\fboxsep}{0pt}
		\colorbox{lightgray}{%
			\begin{tabular}{|l|}
				\hline 
				$  \vdots  $
				\\
				4. Construct the matrix $ {\hat B\left( {{q^{ - 1}}} \right)\left( k \right){\phi _r}\left( k \right)} $ and compute residual 
				\\
				error $ \hat \theta _M^T\left( k \right){\phi _r}\left( k \right) $;
				\\
				5. Update $ \hat{\theta}_D(k) $ using Eq.~(\ref{eq:errorNew}) and compute $ u_a(k) $ from Eq.~(\ref{eq:ffwd}).
				\\
				\hline	
			\end{tabular}%
		}\endgroup
		\captionof{table}{Improved Dreict Adaptive Feedforward Control}\label{tab:updatealgorithm}
	\end{center}
\end{spacing}

\section{Experiment Results and Conclusion}\label{sec:experiment}
\begin{figure}[t!]
	\centering
	\includegraphics[width=1\columnwidth]{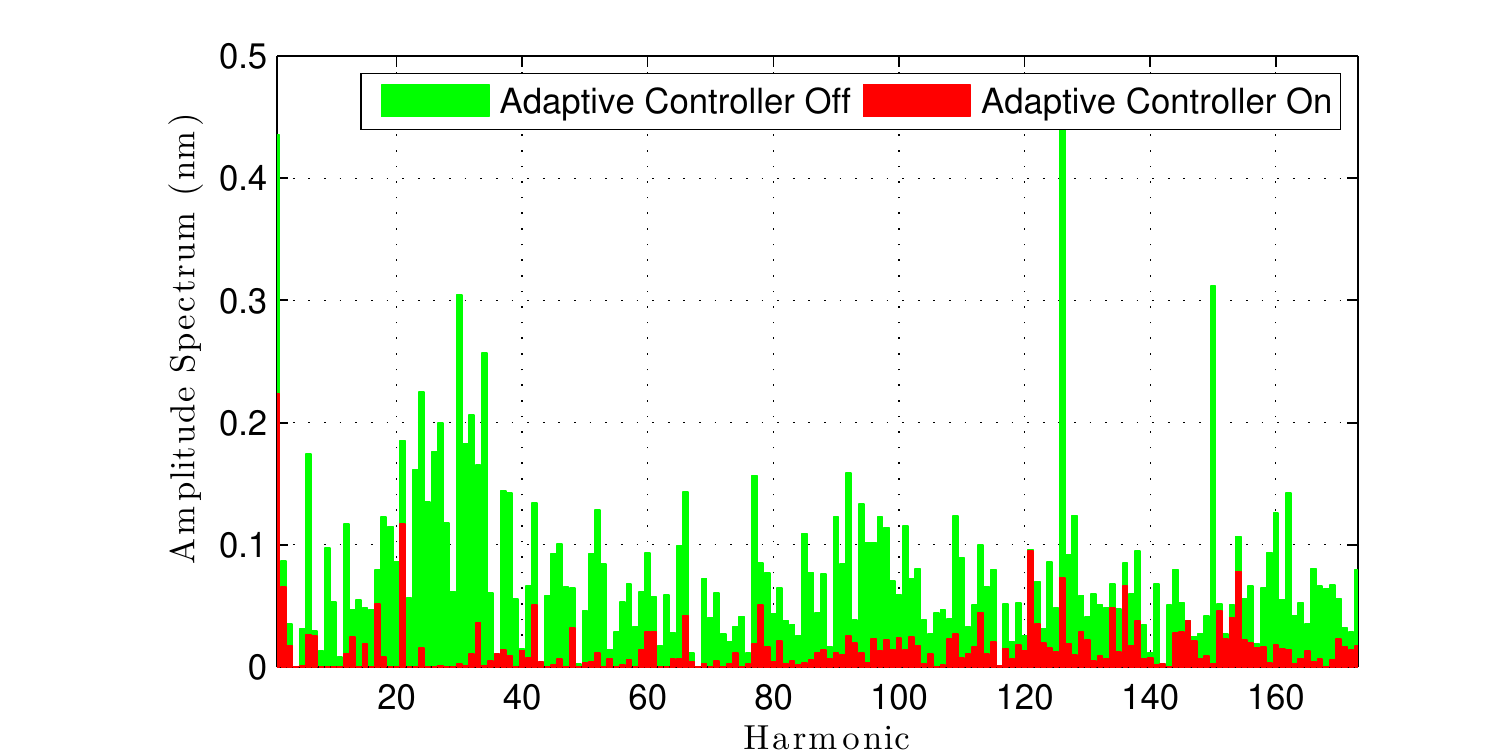}
	\caption{Spectrum comparison.}
	\label{Fig:spectrum}
\end{figure}

\begin{figure}[t!]
	\centering
	\includegraphics[width=1\columnwidth]{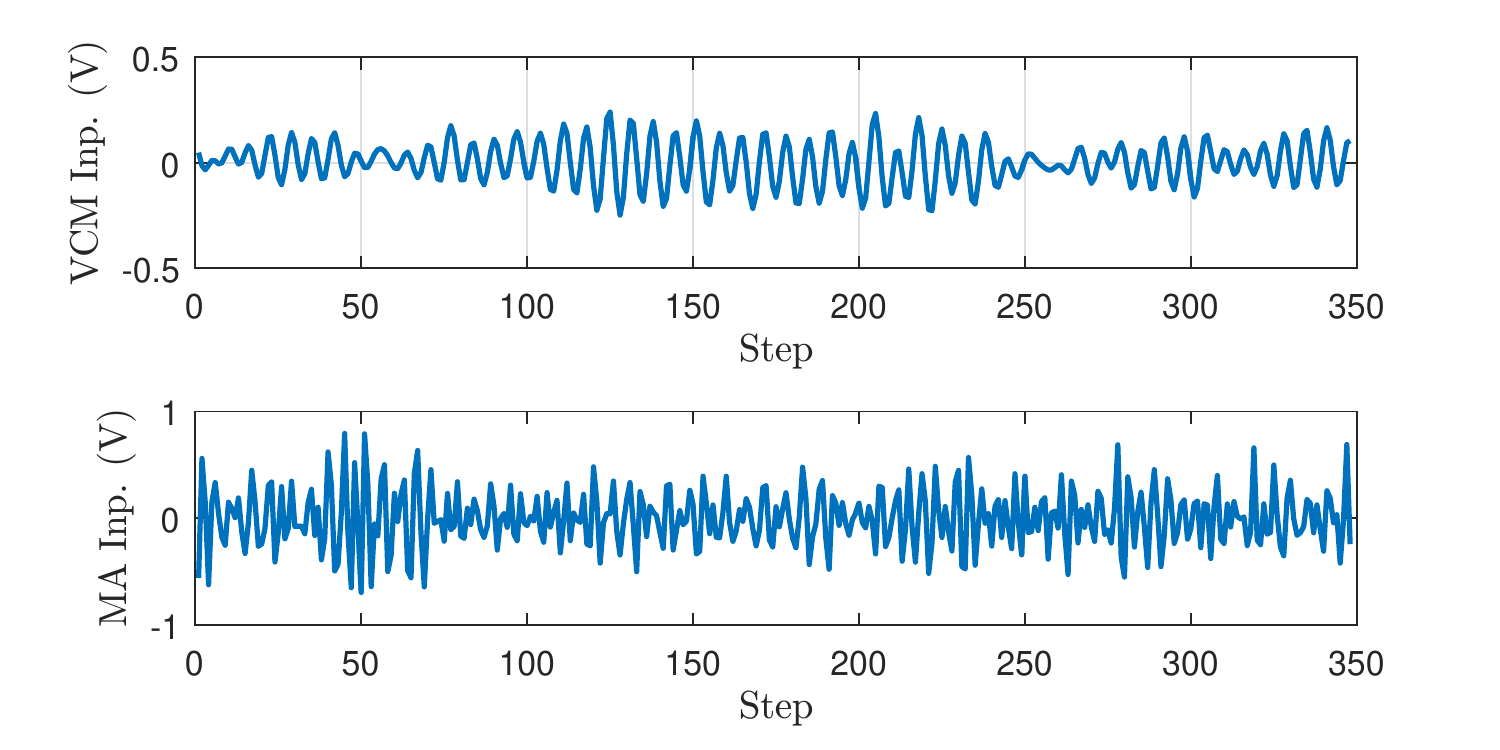}
	\caption{Feedforward signal for VCM and MA.}
	\label{Fig:uffwd}
\end{figure}

We implement both of the two algorithms in MATLAB simulation and the real time experiment setup on dual-stage HDD. In simulation, $ R $ and RRO together with NRRO are modeled from real system measurement data. Since the simulation and experiment results were very close, only experiment results using the improved version are shown in Fig.~\ref{Fig:spectrum}, where RRO is reduced to NRRO level. In simulation as well as in experiments, VCM was responsible for the low frequency RRO (harmonics up to 58), while MA was responsible for the high frequency RRO (harmonics from 59 to 173). As a result, the feedforward control signal in one disk revolution shown in Fig.~\ref{Fig:uffwd} for the VCM consists of low frequency contents while for the MA it has high frequency components.

\section*{Acknowledgment}
Financial support for this study was provided by a grant from the Advanced Storage Technology Consortium (ASTC). 

\bibliographystyle{asmems4}
\bibliography{0-ISPS2016-DA}

\end{document}